\documentclass[a4paper,11pt]{article}
\usepackage{pos}
\usepackage{slashed,bbold,bm,mathtools,wrapfig,epsfig}

\renewcommand{\logo}{\relax}

\newcommand{\crt}{\\[2mm]}
\newcommand{\nn}{\nonumber}
\newcommand{\beq} {\begin{equation}}
\newcommand{\eeq} {\end{equation}}
\newcommand{\beqa} {\begin{eqnarray}}
\newcommand{\eeqa} {\end{eqnarray}}

\newcommand{\bs}[1]{\boldsymbol{#1}}

\newcommand{\ie}{{\it i.e.}}

\newcommand{\eg}{{\it e.g.}}

\newcommand{\as}{\alpha_s}
\newcommand{\lqcd}{\Lambda_{QCD}}
\newcommand{\la}{\Lambda}

\newcommand{\order}[1]{${\mathcal O}\left(#1 \right)$}

\newcommand{\eq}[1]{(\ref{#1})}

\newcommand{\inv}[1]{\frac{1}{#1}}
\newcommand{\halft}{{\textstyle \frac{1}{2}}}
\newcommand{\quart}{{\textstyle \frac{1}{4}}}

\newcommand{\intt}{{\textstyle \int}}
\newcommand{\ket}[1]{\left\vert{#1}\right\rangle}

\newcommand{\bra}[1]{\langle{#1}\vert}

\newcommand{\comb}[2]{\big[{#1},{#2}\big]}

\newcommand{\acomb}[2]{\big\{{#1},{#2}\big\}}

\newcommand{\mE}{\mathcal{E}}

\newcommand{\mH}{\mathcal{H}}

\newcommand{\mL}{\mathcal{L}}

\newcommand{\xv}{{\bs{x}}}

\newcommand{\yv}{{\bs{y}}}
\newcommand{\zv}{{\bs{z}}}

\newcommand{\Av}{{\bs{A}}}

\newcommand{\Ev}{{\bs{E}}}

\newcommand{\Pv}{{\bs{P}}}
\newcommand{\Pva}{{\bs{P}_A}}
\newcommand{\Pvb}{{\bs{P}_B}}

\newcommand{\gv}{\bs{\gamma}}

\newcommand{\gz}{\gamma^0}

\newcommand{\gf}{\gamma_5}

\newcommand{\rar}{\rightarrow}

\newcommand{\nv}{\bs{\nabla}}

\newcommand{\rnab}{{\overset{\rar}{\nv}}\strut}

\newcommand{\alv}{{\bs{\alpha}}}

\newcommand{\Phir}{\Phi^{\scriptstyle{(0)}}}
\newcommand{\Phip}{\Phi^{\scriptscriptstyle{(P)}}}

\newcommand{\dt}{\partial_t}

\title{Bound state basics}

\author*{Paul Hoyer}

\affiliation{Department of Physics, University of Helsinki\\
  POB 64, FIN-00014 University of Helsinki, Finland}

\emailAdd{paul.hoyer@helsinki.fi}

\abstract{
Perturbative expansions for atoms in QED are developed around interacting states, typically defined by the Schr\"odinger equation. Calculations are nevertheless done using the standard  Feynman diagram expansion around free states \cite{Adkins:2022omi}.
The classical $-\alpha/r$ potential is then obtained through an infinite sum of ladder diagrams. The complexity of this approach may have contributed to bound states being omitted from QFT textbooks, restricting the field to select experts.\\[2mm]
The confinement scale $\sim$ 1 fm of QCD must be introduced without changing the Lagrangian. This can be done via a boundary condition on the gauge field, which affects the bound state potential. The absence of confinement in Feynman diagrams may be due to the free field boundary condition.\\[2mm]
Poincar\'e invariance is realized dynamically for bound states, \ie, the interactions are frame dependent. Gauge theories have instantaneous interactions, due to gauge fixing at all points of space at the same time. In bound state perturbation theory each order must have exact Poincar\'e invariance. This is non-trivial even for atoms at lowest order \cite{Jarvinen:2004pi}.\\[2mm]
I summarize a perturbative approach to equal time bound states in QED and QCD, using a Fock expansion in temporal ($A^0=0$) gauge \cite{Hoyer:2021adf}. The longitudinal electric field $\Ev_L$ is instantaneous and need not vanish at spatial infinity for the constituents of color singlet states in QCD. Poincar\'e covariance determines the boundary condition for $\Ev_L$ up to a universal scale, characterized by the gluon field energy density of the vacuum. A non-vanishing density contributes a linear term to the $q\bar q$ potential, while $qqq,\ q\bar qg$ and $gg$ color singlet states get analogous confining potentials.
}

\FullConference{42nd International Conference on High Energy Physics (ICHEP2024)\\
18-24 July 2024\\
Prague, Czech Republic\\}

\begin{document}
\maketitle

\section{Bound state perturbation theory} \label{secI}

Perturbation theory for atoms is essentially different from the perturbative expansion of scattering amplitudes  \cite{Adkins:2022omi}.
Bound states are expanded around an interacting state, \eg, defined by the Schr\"odinger equation. The initial state wave function is non-polynomial (typically exponential) in $\alpha$, giving the expansion non-perturbative features. This contrasts with Feynman diagrams, which expand around free states and lack bound state poles.

Confinement causes hadrons to be strongly bound, which may affect the QCD coupling $\as$. Gribov argued that the running of $\as(Q^2\to 0)$ stops at $\as^{crit}= \pi C_F^{-1}(1-\sqrt{2/3}) \simeq 0.43$ \cite{Dokshitzer:2004ie}. This is close to its value in the ``Cornell potential'',
\begin{align} \label{mo3}
V(r) = V'r-\frac{4}{3}\frac{\as}{r} \ \ \ \text{where}\ \ V' \simeq 0.18\ \text{GeV}^2, \ \ \as \simeq 0.39
\end{align}
which was determined by fits to charmonia and later confirmed by lattice QCD \cite{Eichten:2007qx}. 
The confining potential $V'r$ is not needed in quarkonium transitions and decays, which are evaluated perturbatively as for atoms. 

Even strongly bound hadrons can be described as $q\bar q$ and $qqq$ states. The confining force appears not to create additional quarks and gluons. The valence quarks must be bound by an \textit{instantaneous} potential since gluons propagate with finite speed, contributing higher Fock states. Instantaneous interactions are possible in gauge theories due to gauge fixing in all of space at an instant of time. 

The Lagrangians of QED and QCD lack the $\dt A^0$ and $\nv\cdot\Av$ derivatives of the gauge field. Hence $A^0$ and $\Av_L$ do not propagate in space-time.  In Coulomb gauge the longitudinal field $\Av_L$ is restricted by $\nv\cdot\Av(t,\xv)=0$, while Gauss' law, $\delta S_{QCD}/\delta A^0=0$, expresses $A^0$ in terms of the propagating fields. Coulomb gauge is commonly used in bound state calculations. However, quantization requires constraints since the field conjugate to $A^0$ is missing, $\delta S_{QCD}/\delta (\dt A^0)=0$. 

I consider temporal gauge ($A^0(t,\xv)=0$) since canonical quantization is straightforward. The electric field $\Ev=-\dt\Av$ is conjugate to the gluon field $\Av$ \cite{Leibbrandt:1987qv}. Physical states $\ket{\mathrm{phys},t}$ are required to be invariant under time independent gauge transformations. In QCD this condition is
\begin{align} \label{eII29}
\nv\cdot \Ev^{a}(t,\xv)\ket{\mathrm{phys},t} = g\big[- f_{abc}\Av_b\cdot \Ev_c+q^\dag T^aq\big]\ket{\mathrm{phys},t} \equiv g\mE_a(t,\xv)\ket{\mathrm{phys},t}
\end{align}
Hadrons are expanded in quark and gluon Fock states, schematically for a meson
\begin{align} \label{ich0}
\ket{\mathrm{meson},t}= \Phi_{q\bar q}\,q\bar q\ket{0}+\Phi_{q\bar qg}\,q\bar qg\ket{0}+\ldots
\end{align}
The operator $\mE_a(t,\xv)$ of \eq{eII29} acts for each Fock component on the gauge dependent quark and gluon fields $q\bar q,\ q\bar qg,\ldots$, but annihilates the gauge invariant vacuum $\ket{0}$. This associates to each Fock state in \eq{ich0} an instantaneous, longitudinal electric field $\Ev_L^a(t,\xv)$ which induces interactions (binding) through the Hamiltonian,
\begin{align} \label{eII25}
\mH_{QCD} &= \int d\xv\big[\halft \Ev_a\cdot \Ev_a +\quart F_a^{ij}F_a^{ij}+q^\dag(-i\alv\cdot\nv+m\gz-g\,\alv\cdot \Av_a T^a)q\big]
\end{align}
 
A perturbative expansion for bound states in temporal gauge is now apparent. \textit{E.g.,} for Positronium $\ket{\mathrm{Pos},t}$ we may expand around the lowest Fock state $\Phi_{e\bar e}\,e\bar e\ket{0}$. Eq.~\eq{eII29} determines the classical electric field $\Ev_L$ for each position of the electron and positron in $e\bar e\ket{0}$. The bound state condition $\mH_{QED}(t)\ket{\mathrm{Pos},t} = (2m_e+E_B)\ket{\mathrm{Pos},t}$ imposes (in the non-relativistic limit of the rest frame) the Schr\"odinger equation on the wave function $\Phi_{e\bar e}$. The interaction terms in $\mH_{QED}$ create, operating on $e\bar e\ket{0}$, also higher Fock states such as $e\bar e\gamma\ket{0}$, weighted by the charge $e$. Determining the binding energy $E_B$ up to a given power of $e$ requires to include Fock states with a corresponding number of constituents. 

This bound state perturbative expansion does not rely on Feynman diagrams, allowing to choose the  boundary condition in solving for $\Ev_L$ from \eq{eII29}. It was checked to give the correct \order{\alpha^4} contribution to Positronium hyperfine splitting, which arises from transverse photon exchange and the annihilation channel \cite{Hoyer:2021adf}. It also agrees with the momentum dependence of atomic wave functions found in \cite{Jarvinen:2004pi}. Further checks are obviously desirable.

\section{Color confinement from a boundary condition} \label{secII}

The introduction of the confinement scale $\lqcd \sim 1$ fm$^{-1}$ is a central issue in any approach to QCD. This scale does not appear in the Lagrangian $\mL_{QCD}$ nor in the equations of motion. Renormalization brings a scale via ``dimensional transmutation'', but this does not explain confinement. The QED coupling $\alpha$ also runs, and the scale is given by the electron mass. In lattice QCD the size of the lattice in fermi is set by fixing the mass of a hadron in GeV.

In the present framework the only possibility to introduce $\lqcd$ (without changing $\mL_{QCD}$) is through a boundary condition on the solution of \eq{eII29} for the longitudinal electric field $\Ev_L$. In QED one requires $\lim_{|\xv| \to\infty}\Ev_L(\xv)=0$ to avoid long-range forces. The situation in QCD is different. A (globally) color singlet state such as $\sum_C |q^C(\xv_1)\bar q^C(\xv_2)\rangle$ does not generate a color octet field $\Ev_L^a(\xv)$. Color symmetry ensures that $\Ev_L^a$ cancels in the sum over the quark colors $C$. Hence $\Ev_L^a =0$ for an external observer of a color singlet bound state.

The quark $q^C(\xv_1)$ of a specific color $C$ interacts with the field of its anti-quark $\bar q^C(\xv_2)$ partner in the Fock state. It is not necessary to require the $\Ev_L^a$ of the $|q^C(\xv_1)\bar q^C(\xv_2)\rangle$ state to vanish at large $|\xv|$ \textit{for each color} $C$, since it cancels in the sum over $C$. We may consider adding a homogeneous solution to $\Ev_L^a$, 
\begin{align} \label{eII31}
\Ev_{L}^a(t,\xv)\ket{\mathrm{phys},t} &= -\nv_x \int d\yv \Big[\kappa\,\xv\cdot\yv + \frac{g}{4\pi|\xv-\yv|}\Big]\mE_a(t,\yv) \ket{\mathrm{phys},t}
\end{align}
where $\mE_a$ is defined in \eq{eII29}. Up to its normalization $\kappa$ (which may depend on $\ket{\mathrm{phys},t}$) the form of the homogeneous $(\nv\cdot \Ev_{L}^a=0)$ solution is mandated by translation and rotation invariance. The contribution of this $\Ev_L^a$ to the Hamiltonian \eq{eII25} is
\begin{align} \label{eII32}
\mH_V &\equiv \halft\int d\xv\,\Ev^a_{L}\cdot \Ev^a_{L} = \int d\yv d\zv\Big\{\,\yv\cdot\zv \Big[\halft\kappa^2\intt d\xv + g\kappa\Big] + \halft \frac{\as}{|\yv-\zv|}\Big\}\mE_a(\yv)\mE_a(\zv)
\end{align} 
The term of \order{\kappa^2} is $\ \propto \int d\xv$, signaling a spatially constant field energy density. It is irrelevant only provided it is universal, \ie, the same for all Fock components of all bound states. This determines the normalization $\kappa$ in \eq{eII31} for each Fock state $\ket{\mathrm{phys},t}$ up to a common scale $\la$.

For meson Fock states $\ket{q\bar q} \equiv  \sum_C\bar q^C(\xv_1)q^C(\xv_2)\ket{0}$ the instantaneous interaction $\mH_V$ is
\begin{align} \label{qcd5}
\mH_V\ket{q\bar q} = \Big[\big(\halft\kappa^2\intt d\xv + g\kappa\big)C_F\,(\xv_1-\xv_2)^2-C_F\frac{\as}{|\xv_1-\xv_2|}\Big]\ket{q\bar q}
\end{align}
where $C_F=4/3$ for color SU(3). The term $\ \propto \int d\xv$ must be independent of $\xv_1-\xv_2$, hence $\kappa \propto 1/|\xv_1-\xv_2|$. The \order{g\kappa} term then gives a linear potential. Subtracting the (infinite but universal) field energy $E_\la\int d\xv$ with $E_\la \equiv\la^4/(2g^2C_F)$,
\begin{align} \label{qcd9}
\mH_V\ket{q\bar q} &=\Big[\la^2|\xv_1-\xv_2|-C_F\frac{\as}{|\xv_1-\xv_2|}\Big]\ket{q\bar q}
\equiv V_{q\bar q}(\xv_1,\xv_2)\ket{q\bar q}
\end{align}
This agrees with the Cornell potential \eq{mo3}, but here it is not restricted to non-relativistic quarks. The hadronic scale $V'=\la^2$ is of \order{\as^0} since it arises from a boundary condition. $\mH_V$ does not create quark pairs or gluons since $\Ev_L^a$ is determined by gauge fixing \eq{eII31}, which does not affect the vacuum $\ket{0}$. This explains why strongly bound hadrons can be dominated by their valence quark Fock states.

The corresponding potentials for other states are readily determined. For baryon Fock states $\ket{qqq} \equiv \epsilon_{ABC}\,q_A^\dag(\xv_1)q_B^\dag(\xv_2)q_C^\dag(\xv_3)\ket{0}$,
\begin{align} \label{ich2}
\mH_V\ket{qqq} = \bigg[&\frac{\la^2}{\sqrt{2}}\sqrt{(\xv_1-\xv_2)^2+(\xv_2-\xv_3)^2+(\xv_3-\xv_1)^2} \nn\crt
&-\frac{2\as}{3}\Big(\inv{|\xv_1-\xv_2|}+\inv{|\xv_2-\xv_3|}+\inv{|\xv_3-\xv_1|}\Big)\bigg]\ket{qqq} \equiv V_{qqq}(\xv_1,\xv_2,\xv_3)\ket{qqq}
\end{align}
When two quarks coincide the baryon potential reduces to the meson one (up to a self-energy), $V_{qqq}(\xv_1,\xv_2,\xv_2) = V_{q\bar q}(\xv_1,\xv_2)$.
The potentials of the $\ket{q\bar q\,g}$ and $\ket{gg}$ Fock states are given in \cite{Hoyer:2021adf}.

\section{Bound states at \order{\as^0}} \label{secIII}

A meson state with CM momentum $\Pv$ at $t=0$ may at \order{\as^0} be expressed as superposition of the above $\ket{q\bar q}$ Fock states. For SU($N_c$) gauge symmetry,
\begin{align} \label{qcd97}
\ket{M,\Pv} = \inv{\sqrt{N_c}}\int d\xv_1 d\xv_2\,\bar q_{\alpha,A}(\xv_1)e^{i\Pv\cdot(\xv_1+\xv_2)/2}\delta^{AB}\Phip_{\alpha\beta}(\xv_1-\xv_2)q_{\beta,B}(\xv_2)\ket{0}
\end{align}
A sum over the repeated Dirac ($\alpha, \beta$) and color ($A, B$) indices is understood. I consider only the linear, \order{\as^0} part of the $V_{q\bar q}$ potential in \eq{qcd9} and neglect the \order{g} interaction term of $\mH_{QCD}$ in \eq{eII25}. Adding the free quark Hamiltonian $\mH_0 = \int d\xv\,q_A^\dag(\xv)(-i\alv\cdot\nv+m\gz)q_A(\xv)$ the eigenstate condition imposes a bound state equation (BSE) on the $4\times 4$ wave function $\Phip_{\alpha\beta}(\xv_1-\xv_2)$,
\begin{align} \label{ich3}
(\mH_0&+\mH_V)\ket{M,P} = E\ket{M,P} \hspace{2cm} E = \sqrt{M^2+\Pv^2} \hspace{2cm} (\mH_0+\mH_V)\ket{0} =0 \nn\crt
&i\nv\cdot\acomb{\alv}{\Phip(\xv)}-\halft \Pv\cdot \comb{\alv}{\Phip(\xv)}+m\comb{\gz}{\Phip(\xv)} = \big[E-V(\xv)\big]\Phip(\xv)
\end{align}
where $\alv=\gz\gv$ and $V(\xv)=\la^2|\xv|$. 

The rest frame ($\Pv=0$) wave functions $\Phir(\xv)$ are found by separating the angular and radial dependence. Even though the BSE is fully relativistic, the allowed meson quantum numbers $J^{PC}$ turn out to agree with those of the non-relativistic quark model. States with spin $J$, $J^z=\lambda$, parity $\eta_P=(-1)^{J+1}$ and charge conjugation $\eta_C=(-1)^{J}$ have wave functions
\begin{align}\label{qcd60}
&\Phir(\xv) = \Big[\frac{2}{M-V}(i\alv\cdot\rnab+m\gz)+1\Big]\gf\,F(r)Y_{J\lambda}(\hat\xv)
\nn\crt
&F''+\Big(\frac{2}{r}+\frac{V'}{M-V}\Big)F' + \Big[\quart (M-V)^2-m^2-\frac{J(J+1)}{r^2}\Big]F = 0 
\end{align}
where $Y_{J\lambda}$ are the standard spherical harmonics. Requiring the radial wave function $F(r)$ to be regular at $V'r=M$ determines the bound state masses $M$. At small quark masses $m$ the states lie on nearly linear Regge trajectories ($J \propto M^2$), with parallel daughter trajectories, as in dual models.

For $\Pv \neq 0$ the variables cannot be separated so the BSE is more difficult to solve. It is far from evident that \eq{ich3} respects boost covariance, with the energy eigenvalue $E=\sqrt{M^2+\Pv^2}$. The homogeneous solution in \eq{eII31} was constrained by translation and rotation invariance, whereas boost invariance is realized dynamically. Boost covariance may be expected only because the action is Poincar\'e invariant and the confinement scale was introduced through a boundary condition, maintaining the QCD equations of motion.

In $D=1+1$ dimensions the boost generator is known for a linear potential \cite{Dietrich:2012iy}. The boosted wave function satisfies \eq{ich3} for all $P$. Up to a spin rotation, the wave function is frame independent when expressed in terms of the square of the kinetic 2-momentum $(E-V,P)$, 
\begin{align} \label{ich4}
\Phip(\tau) = e^{-\sigma_1\zeta/2}\Phir(\tau)e^{\sigma_1\zeta/2} \hspace{1cm} \tanh\zeta = \frac{P}{E-V} \hspace{1cm} V'\tau \equiv (E-V)^2-P^2
\end{align}
where $\sigma_1$ is the Pauli matrix representation of the Dirac matrix $\gz\gamma^1$. When $E \gg V'|x|$ this corresponds to standard Lorentz contraction, $V'\tau = M^2-2EV+V^2 \simeq M^2-2V'E|x|$ and $\zeta \simeq \xi$ with $\tanh\xi=P/E$.

A relation similar to \eq{ich4} holds in $D=3+1$ dimensions when the potential is linear and $\xv$ is aligned with $\Pv$. The $\Pv$-dependence implied by \eq{ich3} is not known analytically for general values of $\xv$. However, the (transition) electromagnetic form factor 
\begin{align}  \label{f2}
F_{AB}^\mu(y) &= \bra{M_B, \Pvb}j^\mu(y)\ket{M_A,\Pva} \hspace{2cm} j^\mu(y) = \bar q(y)\gamma^\mu q(y)
\end{align}
was shown to be gauge invariant and to transform as a four-vector \cite{Hoyer:2023noz}. These properties are non-trivial and support the validity of exact Poincar\'e covariance also in $D=3+1$ dimensions.

The states \eq{qcd97} are mutually orthogonal, $\bra{M_B,\Pv_B}M_A,\Pv_A\rangle \propto \delta^3(\Pv_A-\Pv_B)\delta_{A,B}$. However, there is a non-vanishing overlap between one- and multi-hadron states, \eg,
\begin{align} \label{ich5}
\bra{M_B,\Pv_B;M_C,\Pv_C}M_A,\Pv_A\rangle = \bra{\Phi_B^{(P_B)};\Phi_C^{(P_C)}}\Phi_A^{(P_A)}\rangle\,(2\pi^3)\delta^3(\Pv_A-\Pv_B-\Pv_C)/\sqrt{N_c}
\end{align}
where $\bra{\Phi_B^{(P_B)};\Phi_C^{(P_C)}}\Phi_A^{(P_A)}\rangle$ is given by the wave functions. The overlap does not describe a decay $A \to B+C$, since no interaction is involved. It is rather related to the duality of hadron physics, where a state $A$ is found to describe that of $B+C$ in an average sense. For example, in $e^+e^-\to$ hadrons the initial $q\bar q$ state gives a rough description of the final hadron state \cite{Melnitchouk:2005zr}. The energy of a state defined at an instant of time is indeterminate. The wave functions $\Phip(\xv)$ which solve \eq{ich3} are not normalizable, apparently due to the overlap with states of arbitrarily high energy. Physical quantities involve an average over interaction times, removing the contributions of short-lived states.

\section{Summary} \label{secIV}

The approach to equal-time bound states outlined above differs in several respects from previous methods. Temporal $(A^0=0)$ gauge is used because it preserves translation and rotation symmetry, and allows quantization without constraints. An instantaneous electric field is associated with charged constituents in states that are invariant under time independent gauge transformations. This field provides binding without particle creation. 

The perturbative expansion is based on Fock states, keeping all constributions to the binding energy up to a given power of the coupling constant. This should ensure the cancellation of infrared singularities between different Fock states. The Fock state constituents interact through their instantaneous electric field implied by temporal gauge.

The QCD scale $\la_{QCD}$ is introduced through a boundary condition on the solution for the color electric field generated by the quark and gluon color charges. Translation and rotation symmetry determines the form of the homogeneous solution, up to its normalization. This gives rise to a scale which is common to all Fock states and characterized by the gluon field energy density of the vacuum. The boundary condition introduces a confining potential to the Hamiltonian. For color singlet states the electric field cancels in the sum over quark and gluon colors, giving no effects for observers external to the bound state.

Boost covariance is dynamical (interaction dependent) for states defined at equal time, and provides a stringent test of the approach. It is verified explicitly in $D=1+1$ dimensions, and checked for electromagnetic form factors in $D=3+1$. Several aspects of the dynamics relate to the conspicuous but poorly understood features of duality in hadron physics.

\bibliographystyle{JHEP}
\bibliography{refs.bib}

\end{document}